\documentclass[aps,prl,twocolumn,groupedaddress,longbibliography,notitlepage, superscriptaddress]{revtex4-1}
\usepackage{amsmath}
\usepackage{graphicx}
\usepackage{amssymb}
\usepackage{xfrac}
\usepackage{amsmath}
\usepackage{braket}
\usepackage[utf8]{inputenc}
\usepackage{xcolor}
\DeclareUnicodeCharacter{0449}{w}

\newcommand{\beginsupplement}{%
        \setcounter{table}{0}
        \renewcommand{\thetable}{S\arabic{table}}%
        \setcounter{figure}{0}
        \renewcommand{\thefigure}{S\arabic{figure}}%
     }

\begin{document}

\title{Accelerating ultrafast spectroscopy with compressive sensing}

\author{Sushovit Adhikari}
\affiliation{Center for Nanoscale Materials,\\ Argonne National Laboratory, Lemont, Illinois 60439, USA}

\author{Cristian L. Cortes}
\affiliation{Center for Nanoscale Materials,\\ Argonne National Laboratory, Lemont, Illinois 60439, USA}

\author{Xiewen Wen}
\affiliation{Center for Nanoscale Materials,\\ Argonne National Laboratory, Lemont, Illinois 60439, USA}

\author{Shobhana Panuganti}
\affiliation{Department of Chemistry, \\ Northwestern University, 2145 Sheridan Road, Evanston,  Illinois 60208}

\author{David J. Gosztola}
\affiliation{Center for Nanoscale Materials,\\ Argonne National Laboratory, Lemont, Illinois 60439, USA}

\author{Richard D. Schaller}
\affiliation{Center for Nanoscale Materials,\\ Argonne National Laboratory, Lemont, Illinois 60439, USA}
\affiliation{Department of Chemistry, \\ Northwestern University, 2145 Sheridan Road, Evanston,  Illinois 60208}

\author{Gary P. Wiederrecht}
\affiliation{Center for Nanoscale Materials,\\ Argonne National Laboratory, Lemont, Illinois 60439, USA}

\author{Stephen K. Gray}
\email[]{gray@anl.gov}
\affiliation{Center for Nanoscale Materials,\\ Argonne National Laboratory, Lemont, Illinois 60439, USA}

\begin{abstract}

% Ultrafast spectroscopy is an important tool for studying photoinduced dynamical processes in atoms, molecules and nanostructures. Typically, the time to perform these experiments ranges from several minutes to hours depending on the choice of spectroscopic method. For example, while ultrafast transient absorption spectroscopy takes several minutes, terahertz spectroscopy takes hours. It is desirable to reduce the experimental time overhead not only to shorten time and other laboratory resources, but also to make it possible to examine fragile specimens which degrade during long experiments. In this article, we motivate using compressive sensing to significantly shorten the experimental time overhead by reducing the total number of measurements taken during data acquisition. We apply this technique to experimental data from ultrafast transient absorption spectroscopy and ultrafast terahertz spectroscopy and show that good estimates can be obtained with as low as 15\% of the total measurements, implying a shortening in experimental time by at least 6-fold. 

Ultrafast spectroscopy is an important tool for studying photoinduced dynamical processes in atoms, molecules, and nanostructures. Typically, the time to perform these experiments ranges from several minutes to hours depending on the choice of spectroscopic method. It is desirable to reduce this time overhead to not only to shorten time and laboratory resources, but also to make it possible to examine fragile specimens which quickly degrade during long experiments. In this article, we motivate using compressive sensing to significantly shorten data acquisition time by reducing the total number of measurements in ultrafast spectroscopy. We apply this technique to experimental data from ultrafast transient absorption spectroscopy and ultrafast terahertz spectroscopy and show that good estimates can be obtained with as low as 15\% of the total measurements, implying a 6-fold reduction in data acquisition time.

\end{abstract}

\maketitle

\section{Introduction}

Ultrafast spectroscopy has found a wide range of applications to study time-resolved ultrafast dynamical processes \cite{ultrafastReview, UltrafastEx1, UltrafastEx2, UltrafastEx4, UltrafastEx5}. 
Many techniques have been developed spanning different time and photon energy ranges, including ultrafast transient absorption spectroscopy, time-resolved photoelectron spectroscopy, multidimensional spectroscopy, and terahertz spectroscopy \cite{UltrafastEx3, weiner2011ultrafast}. 
These techniques can be very time consuming with acquisition times varying drastically depending on the method. 
Reducing the time overhead is important, not only for efficiency, but also for making it possible to examine specimens which degrade quickly due to prolonged exposure to a laser beam.

Here we show how to significantly shorten the duration of ultrafast spectroscopy with compressive sensing. We apply compressive sensing to two important ultrafast techniques: ultrafast transient absorption spectroscopy and ultrafast terahertz spectroscopy. The specimen chosen for the transient absorption is a 50 nm diameter colloidal TiN nanoparticles in water, which are of growing interest as refractory metal nanostructures resistant to heat or optical damage for plasmonics applications \cite{guler2015}. This specimen was chosen due to its high degree of optical scattering which makes it extremely challenging and very time consuming to acquire data with reasonable signal to noise ratio. For our experiment, the data acquisition time was about four hours. For ultrafast terahertz spectroscopy, measurements were performed on a methylammonium lead iodide (MAPbI$_3$) thin film that was spin casted on quartz. This material class is of interest for solar energy conversion and is believed to exhibit long carrier lifetimes owing to low frequency lattice deformations that may help screen charges. The data acquisition for full 2D time-resolved THz experiment took about seven hours to complete, and could require more time for higher signal to noise ratio. These conditions challenge the stability of both laser systems and many specimens. To overcome these difficulties, we show that by taking sparse, random measurements in time, thereby taking a fraction of the total measurements compared to conventional experiments, compressive sensing can faithfully reconstruct the full experimental result.   

\begin{figure} 
    \centering
    \includegraphics[width=8.5cm]{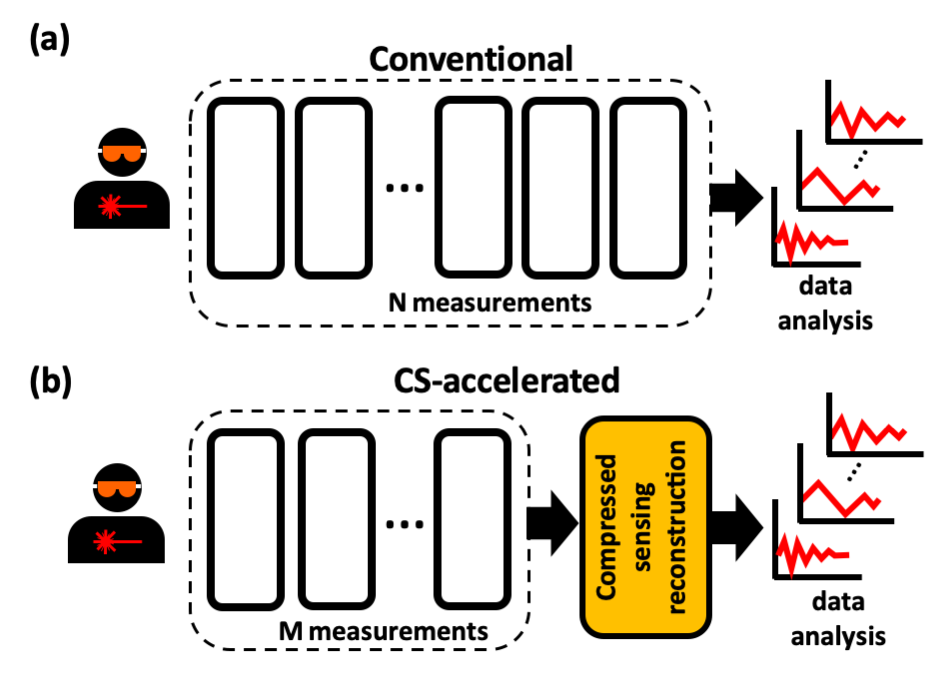}
    \caption{Comparison between  conventional and CS-accelerated experiments. 
    %In a conventional experiment, $N$ measurements are taken. In the CS-accelerated scheme, $M << N$ measurements are taken randomly in time and  fed into a CS-reconstruction algorithm to estimate the full $N$ signal. 
    (a) shows a conventional experiment which requires $N$ measurements for the fully resolved result. (b) shows the CS-accelerated scheme with $M << N$ measurements to yield an estimate of the fully resolved result.}
    \label{fig:General_idea}
\end{figure}

\section{Compressive Sensing for Ultrafast spectroscopy experiments}

Recently, there has been wide interest in using different techniques to speed up optical experiments \cite{cortes2020accelerating, you2020, kudyshev2019rapid, CShaar, Simmerman}. 
Compressive Sensing (CS) is one such technique for efficiently acquiring and reconstructing signals \cite{cs1, cs2, cs3, cs4, cs5}. 
It has successfully been applied in many fields, including magnetic resonance imaging (MRI), fluorescence microscopy, multi-dimensional nuclear magnetic resonance (NMR) spectroscopy, quantum imaging, and quantum tomography \cite{CSMRI, CSFluorescence, CSNMR, CSQO1, CSQO2, CSQO3, CSQO4, CSQO5, CSQO6, xiewen}. CS has also been used in multidimensional spectroscopy for applications in chemistry with impressive speed-ups shown \cite{andrade2012application, sanders2012compressed, dunbar2013accelerated}. We propose compressive sensing for material science and condensed matter physics where a different set of ultrafast spectroscopic methods are used, such as transient absorption and ultrafast terahertz spectroscopy. We hope that our work helps bridge the gap between the CS community and the ultrafast spectroscopy community where advanced algorithmic methods are not commonly used.

Generally, the number of measurements, $N$, to capture full information of a signal is determined by the Nyquist-Shannon sampling theorem: the sampling rate should be at least twice the highest frequency of the signal, 2$f_{max}$ \cite{cover2012elements}.  If some maximum
time $T$ is required to observe the dynamics or infer a spectrum of a given resolution, then $N$ = $T/\Delta t$ = $2f_{max}T$ total measurements would be needed, with $\Delta t$ being the time between measurements (inverse sampling rate). This latter analysis is correct if one only has an upper limit for $f_{max}$ and {\em no} other knowledge about the signal. CS overcomes this limit by invoking a sparsity assumption of the signal in some known basis. When a signal is transformed to this basis, most of the coefficients are negligibly small. The existence of such a basis can be used to significantly reduce the total number of measurements required to reconstruct the full signal. Many natural signals are sparse in the Fourier domain. Since time-domain signals are usually real, the discrete cosine transform (DCT) is widely used for compression and CS reconstruction. Other transformations, such as Haar, total variation (TV) and Hadamard transformations are also widely used \cite{CShaar, CShadamard, CSTV}. 

CS reconstructs a signal by solving the convex optimization problem,
\begin{equation}
    \min_{\tilde{x}}|| \tilde{x} ||_{1} \; \text{subject to} \;  A\tilde{x} = y.
    \label{CS}
\end{equation}
% Here, $x$ is the $N$-dimensional vector and corresponds to the full experimental signal that we want to reconstruct. $\Psi$ is the transform domain in which $x$ is sparse (in our case, we use DCT, Haar and Hadamard transform). $||.||_{1}$ is the $l_{1}$ norm, i.e. the sum of the absolute values of each elements. $y$ is the $M$-dimensional measurement vector obtained by interacting $x$ with $M \times N$ sensing matrix $A$, which is just a subset of identity matrix in our case. The algorithm will search for most sparse $x$ satisfying the above condition.
Here, $\tilde{x} = \psi x$ is the $N \times 1$ sparse solution vector and $\psi$ is the transformation matrix that takes the signal $x$ (e.g., transient absorption), to a sparse basis. We use the DCT, Haar and Hadamard transformations as possible choices for $\psi$. $||.||_{1}$ is the $l_{1}$ norm, i.e., the sum of the absolute values of the components of $\tilde{x}$. 
The most sparse solution is given  by minimizing number of nonzero components of the solution vector $\tilde{x}$ or $l_{0}$ norm. However, $l_{0}$ minimization is non-convex and falls under \textbf{NP}-\textit{hard} computational complexity which is very difficult to solve. $y$ is the $M \times 1$ vector representing the small number ($M\ll N$) of random measurements taken in the experiment. $A = \phi \psi^{-1}$ is a $M \times N$ matrix, with $\phi$ representing an $M \times N$ random measurement matrix, which we take to be a submatrix of the $N \times N$ identity matrix  with $M$ rows chosen randomly. For a $K$-sparse signal, having $K$ nonzero coefficients, the above optimization problem is able to faithfully reconstruct the signal with approximately $K \log{ \big( \frac{N}{K} \big)}$ measurements with high probability. Remarkably, it has been shown that no reconstruction algorithm can reconstruct the signal with substantially fewer measurements \cite{cs3}. 
Additional details of this algorithm are specified  in the methods section. 

To see why the CS is well suited to ultrafast spectroscopy,  consider a pump-probe framework, typical of many such experiments. A short pump pulse centered at time $t_0$ excites a specimen and a probe pulse at various later times $t$ = $t_0+\tau$ is used to measure
the evolution of some material response, $\mathcal{R}$ (e.g., absorbance or transmittance) \cite{weiner2011ultrafast}:  
\begin{equation}
    \mathcal{R}(\tau ) = \mathcal{R}_{-\infty} + \Delta \mathcal{R}(\tau ),
\end{equation} 
where $\mathcal{R}_{-\infty}$ is the material response prior to the pump. 
Many measurements are taken at various probe time delays $\tau$, giving information about the full dynamics. 
Such ultrafast experiments can be time consuming due to the need of making repeated measurements with small increments in $\Delta\tau$. CS is ideally suited for ultrafast optics because in a wide variety of material systems, $\mathcal{R}(\tau )$ is often dominated by a small number, or small range, of frequencies, implying the existence of a sparse basis. 
For many systems, the total number of measurements taken in conventional ultrafast  experiments far outweigh the actual number of measurements, $\sim K \log{ \big( \frac{N}{K} \big)}$, required by CS theory. Here, we test the performance of CS signal reconstruction for two protypical experiments: ultrafast transient absorption and ultrafast terahertz spectroscopy.

% \textcolor{black}{
% The duration of ultrafast experiments are long and time consuming due to the need of making measurements over varying time delays. These long durations are mainly due to a need of changing in fine increment or raster scanning of some experimental variable (example, time delays, temperatures, voltages, etc). Instead of raster scanning, one can do random measurements, with the total measurements being a fraction of a raster scan experiment and use CS algorithm to reconstruct the full experimental signal. 
% fraction of random measurements CS can drastically shorten the experiment duration by making a coarse random measurement of these quantities and applying the }

% In terms of detection most pump-probe experiments utilize a slow response time detector compared to the dynamics of interest. The time resolution comes about due to the ability to scan the pump-probe delay precisely.

\begin{figure*}[t!]
    \centering
    \includegraphics[width=18.5cm]{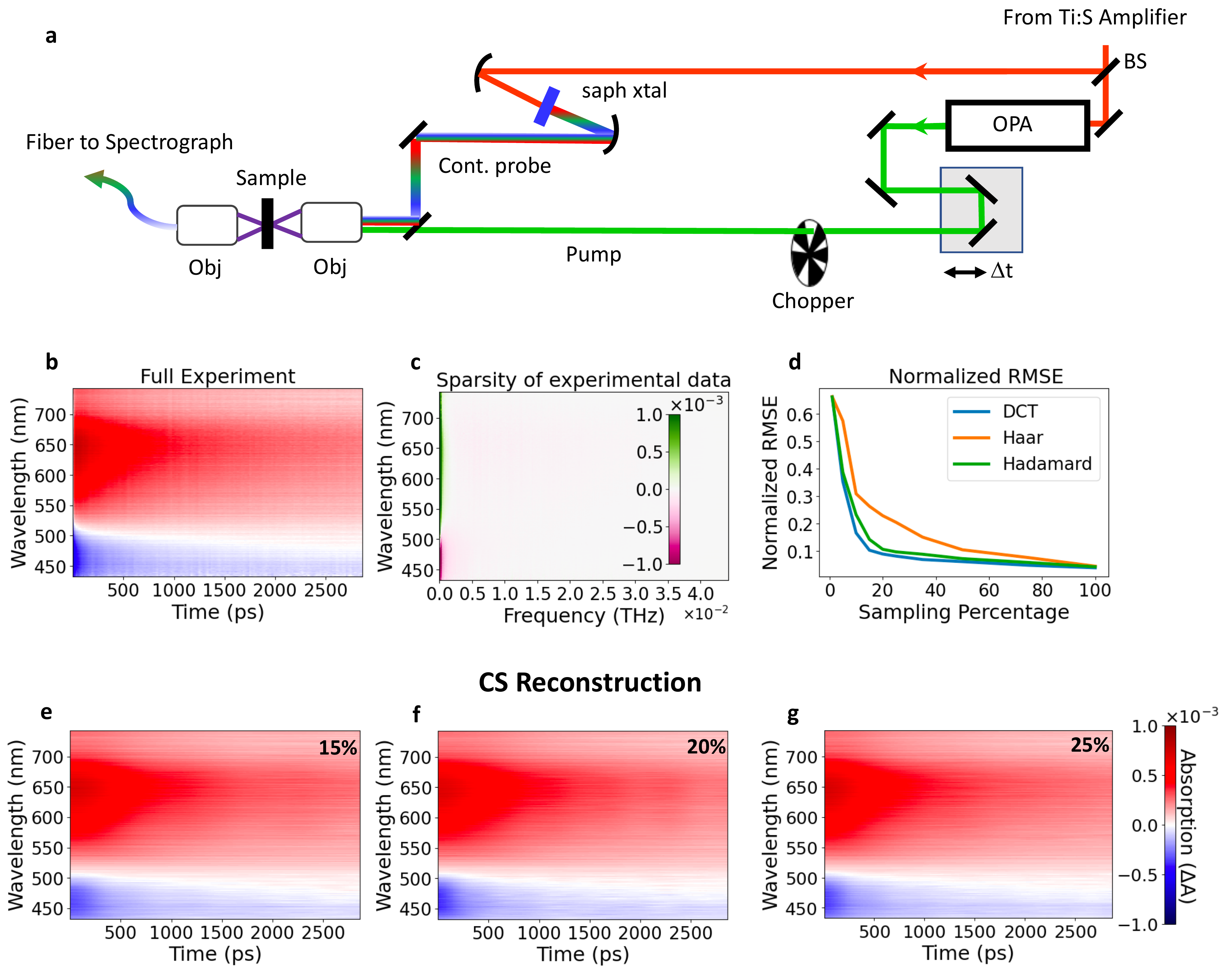}
    \caption{Ultrafast transient absorption spectroscopy. (a) is a schematic of the experimental setup. (b) is the observed change in absorbance across different wavelengths. (c) demonstrates that the full experiment data is very sparse in the DCT domain. Almost all the information is contained in about 5\% of the frequency coefficients. This allows CS to reconstruct signals with fraction of the measurements of a conventional experiment. (d) is the NRMSE for DCT, Haar and Hadamard as a function of sampling percentage.  (e), (f) and (g) shows the CS reconstruction with 15\%, 20\% and 25\% samples, respectively. }
    \label{fig:transient absorption}
\end{figure*}

\begin{figure*}[t!]
    \centering
    \includegraphics[width=18.5cm]{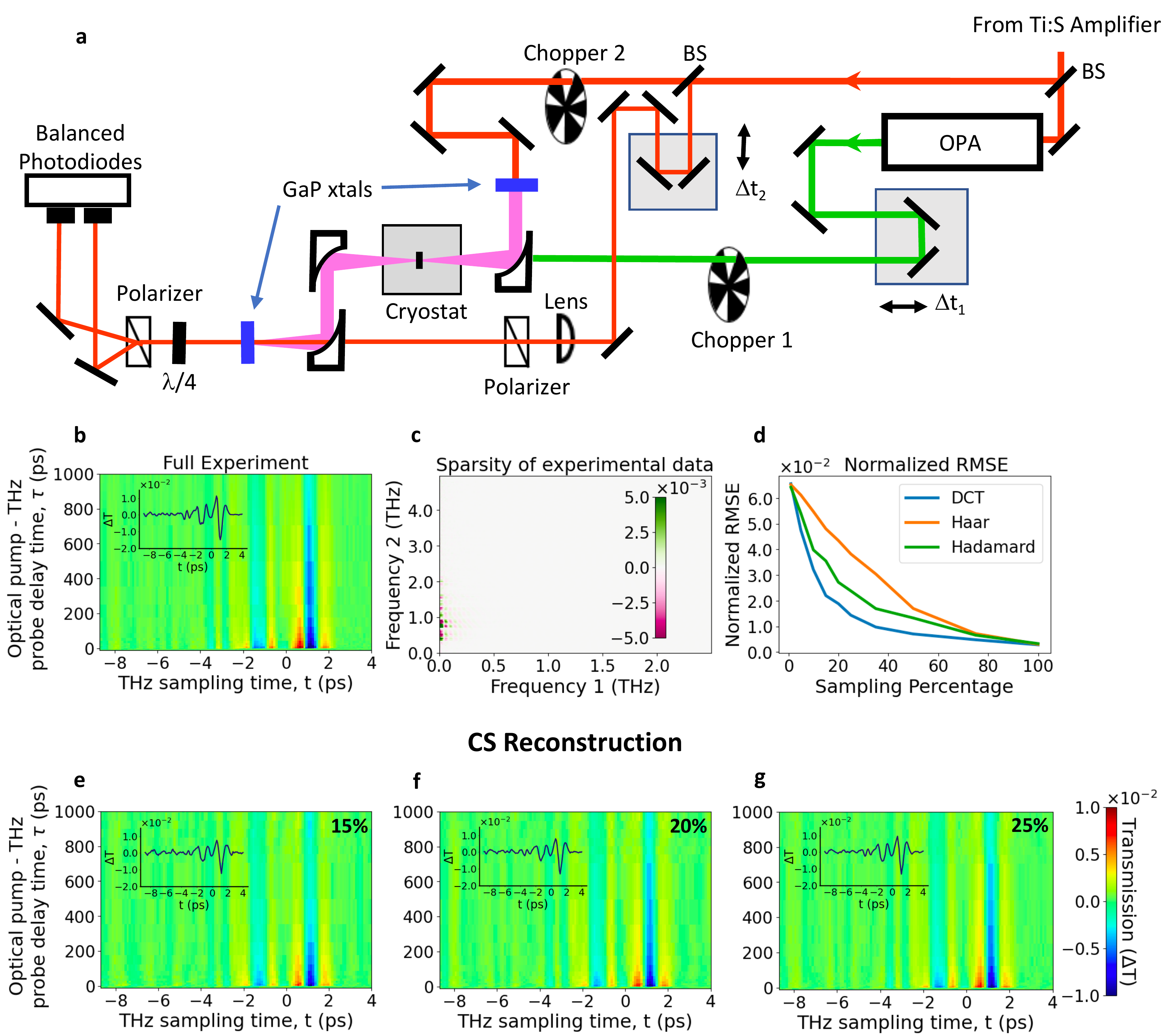}
    
    \caption{Ultrafast terahertz spectroscopy. (a) shows the schematic of the experimental setup. (b) is the full experimental result. (c) shows the sparsity of the signal in DCT domain. Almost all the information is contained in about 7\% of the frequency coefficients. This allows CS to reconstruct the full signal with fraction of the measurements of a conventional experiment. (d) is the comparison of normalized RMSE for DCT, Haar and Hadamard as a function of sampling percentage. (e), (f) and (g) shows the CS reconstruction with 15\%, 20\% and 25\% samples, respectively.   }
    \label{fig:terahertz}
\end{figure*}

\section{Results}

\subsection{Ultrafast Transient Absorption Spectroscopy.}

An ultrafast transient absorption spectroscopy is a pump-probe experiment in which a specimen is excited with a femtosecond pump pulse, followed by a probe pulse with a variable time delay. The change in transmittance or absorbance over various time delays are measured giving information about the properties of the specimen. Transient absorption spectroscopy has been used to characterize an extraordinarily large range of photoinduced dynamical processes ranging from molecular excited states, electronic transitions in nanoparticles, plasmonics, charge separation and transport phenomena, to name just a few \cite{wang2014femtosecond, pendlebury2011dynamics, grancini2011transient}.

A schematic of our ultrafast transient absorption spectroscopy experiment is shown in Figure \ref{fig:transient absorption}-a. The output of an amplified femtosecond laser system (Spectra Physics Tsunami and Spitfire) operating at 5kHz and 800 nm pumps an optical parametric amplifier (OPA) to create wavelength tunable 130 fs pulses. A small portion of the 800 nm amplified light (5\%) is focused into a thin (2mm) sapphire crystal to create a continuum probe. The pump beam is chopped at half the repetition rate to create ``pump-on'' and ``pump-off'' such that a transient absorption signal can be measured with each pump pair. With variable delay of the probe relative to the pump, time-resolved transient absorption spectroscopy can be acquired. The pump and probe beams are focused and spatially overlapped in the specimen (TiN nanoparticles in waters). The probe light is then sent to a spectrograph where the full continuum in the visible spectral range is measured simultaneously for each delay. For this data, each delay required tens of thousands of transient absorption pump-pair measurements. The full data acquisition took four hours due to the high scattering of the colloidal nanoparticle specimen. Next, we show how CS can drastically shorten the duration of experiment. 

% Due to the high degree of optical scattering by the specimen, the experiment is very time consuming and we show how CS can shorten the duration of the experiment.

% In our experiment, we measure the absorbance of TiN nanoparticles in water. The specimen is pumped by a 800 nm laser pulse with a controlled time delay in the picosecond (ps) range and is probed by a continuum white light source. The experiment is repeated 15 times with fixed raster-scanned time delays.  A single experiment provides information about the dynamics over a continuum of wavelengths. Still, the experiment takes several minutes to complete and we show how CS can speedup the duration of experiment.

% \textcolor{orange}{This data required four hours to acquire due to the high scattering of the colloidal nanoparticle specimen.}

Figure \ref{fig:transient absorption}-b is the full experimental data showing the change in absorbance ($\Delta A$) across different wavelengths.  It can be seen that $\Delta A$ varies with wavelength.  The data indicate a strong transient response of the TiN plasmon absorption following photoexcitation. Since the transient absorption measurement is a difference measurement, it is possible to get positive (an absorption with less light transmitted through the specimen) and negative transient absorption signals (typically a bleach of the ground state absorption, with more light transmitted through the specimen), particularly when there is a spectral shift in the excited nanostructure. In Figure 2-b, the data indicate that the plasmon resonance of the TiN nanoparticles, centered at ~520 nm, redshifts upon photoexcitation. For the wavelengths on the lower energy side of 520 nm, more light is absorbed as the peak absorption moves to the red, while the specimen begins to transmit more light at wavelengths on the blue side of 520 nm as the peak absorption moves further to the red away from those wavelengths. Figure \ref{fig:transient absorption}-c shows the sparsity of the full experimental data for each wavelength in DCT domain. We can clearly see that the signal is sparse with almost all of components concentrated near zero frequency. In fact, only about 5\% of the frequency coefficients contain almost all the information about the signal. This suggests that CS can reconstruct the signal with about 15\% of the measurements of a conventional experiment. Figure \ref{fig:transient absorption} (e-g) shows the CS reconstruction at 15\%, 20\% and 25\% respectively. Qualitatively, all these CS reconstruction looks very similar to the full result in Figure \ref{fig:transient absorption}-b. This also corroborates our assumption that about 15\% measurements of a conventional experiment is sufficient to reconstruct the full experimental signal. To quantify the overall signal reconstruction, we compare the normalized root-mean-square-error (NRMSE) for different sampling percentages in Figure \ref{fig:transient absorption}-d. We also compare the DCT with Haar and Hadamard transforms. The DCT performs best among these transformations, followed by Hadamard and Haar transforms. It is interesting to note that the NRMSE drops quickly and does not change much from 15\% to higher sampling percentage. This suggests that the number of measurements as low as 15\% of a conventional experiment is sufficient for a fair reconstruction of the signal.

\subsection{Ultrafast Terahertz Spectroscopy}
Ultrafast Terahertz spectroscopy is another useful technique for investigating specimens with short pulses of terahertz radiation. It is used for examining spectral responses of a specimen in the far infrared and can exhibit sensitivity to pump-induced optical conductivity as well as phonon dynamics in some cases \cite{schmuttenmaer2004exploring, luo2017ultrafast}. Measurement of the terahertz spectral response is often accomplished using electro-optical sampling wherein a time-delayed (often) 800 nm laser pulse is spatially overalapped with the terahertz pulse in a GaP or ZnTe crystal to evaluate the terahertz waveform. As such, ultrafast terahertz spectroscopy can necessitate two dimensions of scanning that increases data acquisition time.

A schematic of ultrafast terahertz spectroscopy is shown in Figure \ref{fig:terahertz}-a. In our experiment, we probe a methylammonium lead iodide (MAPbI3) thin film on quartz at 80 K as a function of a time delayed pump pulse (with time delay $\tau$) of 500 nm. THz probe pulses were produced using via optical rectification in a 300 micron-thick GaP 110 crystal.  Absorption of 500 nm pump photons produces electron-hole pairs in the MAPbI3 specimen that alter the conductivity of the film and change transparency to the THz probe pulse. A recent literature report conveys other physical phenomena in this material including altered optical access to Rydberg states as well as phonon evolution \cite{luo2017ultrafast}.

 Figure \ref{fig:terahertz}-b show the acquired experimental data. Here, the pump-induced change in transmitted THz probe intensity is plotted following optical excitation, where photogenerated carriers alter the sample conductivity owing to the light-induced production of highly polarizable charge carriers and phonon population evolution. In particular, the long carrier lifetimes and diffusion lengths in this material may result from lattice distortions and low frequency vibrational modes, which can be interrogated via this method. Figure \ref{fig:terahertz}-c demonstrates the sparsity of the full experimental result in the DCT domain. We observe that only about 7\% of the coefficients contain almost all the information about the full signal. As before, the fact that the signal is sparse in the DCT domain is the key point which allows us to use compressed sensing for reconstructing the full signal with about 18\% measurements of a conventional experiment. Figure \ref{fig:terahertz} (e-f) shows the CS reconstruction with different percent of measurements (15\%, 20\% and 25\%). These CS reconstructions looks very similar to the full experimental result in Figure \ref{fig:terahertz}b. Again, it verifies our assumption that about 18\% measurements of a conventional experiment reconstructs the full signal. As before, for quantitative assessment, Figure \ref{fig:terahertz}-d shows the NMRSE for DCT, Hadamard and Haar transforms. We observe a similar trend as in the case of ultrafast transient absorption spectroscopy. The DCT performs best among these transformations. Also, the change in NRMSE above 20\% sampling is small, suggesting that a sampling range of 15\% to 25\% gives a good estimate of the full signal.

\section{Discussion}

In conclusion, we applied compressive sensing to experimental data from ultrafast transient absorption spectroscopy and ultrafast terahertz spectroscopy. We showed that CS can faithfully reconstruct the full experiment signal with a fraction of random measurements compared to a conventional experiment. We also compared different transformations and found that DCT to work best in our data.

We envision this technique will be very beneficial in many ultrafast spectroscopy experiments, where data acquisition is time consuming due to raster scanning. For example, in many experiments, a raster scan along temperature or voltage is required. Our method will also provide significant speedup for higher-dimensional ultrafast spectroscopy by reducing the number of measurements in each dimension. Moreover, it should also make is possible to measure fragile specimens which gets degraded on long exposure under the laser.

\section{Methods}

\subsection{Compressive Sensing}
One way of solving Eq. (\ref{CS}) is by formulating it as a ``Lasso'' functional or ``Basis Pursuit DeNoising'' problem as \cite{tibshirani1996, donoho2001}:

% \begin{equation}
%     \underset{x}{\min} \; \frac{1}{2} ||A x - y||^{2}_{2}
%     + \lambda|| \Psi x||_{1}.
% \end{equation}

\begin{equation}
    \underset{\hat{x}}{\min} \; \frac{1}{2} ||A \hat{x} - y||^{2}_{2}
    + \lambda|| \hat{x}||_{1}.
\end{equation}

The above optimization problem can be seen as a trade-off between minimizing the squared error (i.e making $A\hat{x}$ as close to $y$) and finding $\hat{x}$ with a minimal $l1$-norm. Here, $\lambda$ is a regularization parameter and controls the trade-off between sparsity and reconstruction fidelity.  $\lambda$ is data dependent and and has to be estimated for a given data set. One of the method for estimating $\lambda$ is cross-validation which we use and is discussed in detail in supplemental material.

% \subsection{Ultrafast transient absorption spectroscopy} Before applying CS on the experimental data, we preprocess the experimental data. Our initial data consists of 273 different wavelengths ranging from 421 nm to 743 nm and each wavelength has $255$ transmittance coefficients collected over time interval from $t=-5$ ps to $t = 2870$ ps. The 421 nm to 431 nm data is very noisy and we exclude it from analysis \textcolor{orange}{This wavelength range is excluded because of the very small amount of continuum photons generated by 800 nm light incident on the sapphire crystal. Very low levels of light in the probe beam cause large digitization noise in the measured signal.} We also observe a coherent non-resonant response of the solvent at $t=0$ and we exclude this region in subsequent analysis [see supplementary material for details]. After excluding this anomalous region, we interpolate the data such that each wavelength has 256 transmittance coefficients. The interpolation is not necessary for CS reconstruction in general, but we want to compare CS reconstructions with different transformations (DCT, Hadamard and Haar), and Haar and Hadamard transformations require a input signal length of power of two. For CS reconstruction, we randomly sample some percentage of time data and select absorbance coefficients for each wavelength at these selected time. In terms of experiment, this would correspond to fewer measurements at these random times, whereas in a conventional experiment, many more measurements are taken with a fine increment in time delays.

\subsection{Ultrafast transient absorption spectroscopy} Before applying CS, we preprocess the experimental data, which consists of excluding data near $t=0$ due to coherent, non-resonant response, and wavelengths from 421 nm to 431 nm and interpolation \cite{coherentsolvent1,coherentsolvent2} [see supplementary material for details]. The interpolation is not necessary for CS reconstruction in general, but we want to compare CS reconstructions with different transformations (DCT, Hadamard and Haar), and Haar and Hadamard transforms require a input signal length of power of two. For CS reconstruction, we randomly sample some percentage of time data and select absorbance coefficients for each wavelength at these selected time. In terms of experiment, this would correspond to fewer measurements at these random times, whereas in a conventional experiment, many more measurements are taken with a fine increment in time delays.

For quantitative comparison of CS reconstructions at different sampling percentage, we use normalized-root-mean-square-error (NRMSE). The NRMSE is defined as:
\begin{equation}
\textrm{NRMSE} =\frac{1}{M} \sum_{j=1}^{M} \frac{1}{( x_{max}^{j} - x_{min}^{j})}\sqrt{ \frac{ \sum_{i=1}^{N} (\hat{x}^{j}_{i} - x^{j}_{i} )^2 }{N }  }.
\end{equation}

Here, $\hat{x}^{j}$ and ${x}^{j}$ is the CS reconstruction and experimental processed signal for wavelength $j$, $N = 256$ is the length of the signal, $x_{max}^{j}$  and $x_{min}^{j}$ are the maximum and minimum absorption coefficient for wavelength $j$ and $M = 263$ is the number of different wavelengths. The  NRMSE is averaged over 10 runs which corresponds to unique sampled data on each run.

\subsection{Ultrafast Terahertz Spectroscopy}
% Our initial data set consists of 26 and 151  scan along $t1$ and $t2$ respectively. As before, we preprocess the data and interpolate it to make $32 \times 128$ for comparison among DCT, Haar and Hadamard. For the $t1$ data, we interpolate between the last two data points. For the $t2$ axis, we simply cutoff the data where the signal does not have any interesting feature. For the CS reconstruction, we randomly sample from $t1$ and $t2$. This corresponds to making a random coarse measurements in both $t1$ and $t2$ which significantly reduces the duration of experiment. To make it more quantitative, as before we use normalized root-mean-square-error (NRMSE). We define the NRMSE for this case as follows:

As before, we preprocess the data and interpolate it [see supplemental material for details]. For the CS reconstruction, we randomly sample from $\tau$ and $t$. This corresponds to making a random coarse measurements in both $\tau$ and $t$ which significantly reduces the duration of experiment. To make it more quantitative, as before we use normalized root-mean-square-error (NRMSE). We define the NRMSE for this case as follows:

\begin{equation}
    \textrm{NRMSE} = \frac{1}{x_{max} - x_{min}} \sqrt{\frac{\sum_{i,j} (\hat{x}_{i,j} - x_{i,j})^2}  {M N} }.
\end{equation}

Here $\hat{x}$ is the CS reconstruction and $x$ is the full experimental data. $x_{max}$ and $x_{min}$ are the maximum and minimimum of the experimental data. $M=32$ and $N=128$ are the length of the $
\tau$ and $t$ scans respectively. The  NRMSE is averaged over 10 runs which corresponds to unique sampled data on each run as before.

\section{Acknowledgements}

This material is based upon work supported by Laboratory Directed Research and Development (LDRD) funding from Argonne National Laboratory, provided by the Director, Office of Science, of the U.S. Department of Energy under Contract No. DE-AC02-06CH11357. 
Use of the Center for Nanoscale Materials, an Office of Science user facility, was supported by the U.S. Department of Energy, Office of Science, Office of Basic Energy Sciences, under Contract No. DE-AC02-06CH11357.

\bibliography{UltraFastCS.bib}

\onecolumngrid

\section{Supplementary Material}
\beginsupplement

\section{\textbf{Different transformations}}

In the maintext, we used three different transformations (DCT, Haar and Hadamard). Here, we briefly describe how an input signal $x$ of length $N$ is transformed by these transformations.

The DCT transforms a signal $x$ of length $N$ into a signal $X_{N}$ of same length as follows:

\begin{equation}
    X_{k} = \sum_{n=0}^{N-1} x_{n} \cos \bigg[\frac{\pi}{N}  \bigg( n + \frac{1}{2} \bigg) k \bigg] \;\;\; k = 0,...... N - 1.
\end{equation}

In matrix form, the Haar wavelet transform is given by: 
\begin{equation}
W_{N} = 
\begin{bmatrix}
W_{N/2} \otimes [1,1]  \\  
I_{N/2} \otimes [1, 1]
\end{bmatrix}
\end{equation}

where $I_{N/2}$ is the identity matrix of $N/2$-dimension and $\otimes{}$ is the Kronecker product. A signal $x_{N}$ of length $N$ can be transformed as: $X_{N} = W_{N} \times x_{N}$. Similarly, 2-D Haar transform on a $N \times N$ matrix $y_{N}$ gives $Y_{N} = W_{N} \times y_{N} \times W_{N}^{T}$.

\vspace{.3cm}

In matrix form, the Hadamard transformation is defined recursively, with $H_{0}=1$ as:
\begin{equation}
H_{N} = \frac{1} { \sqrt{N} }
 \begin{bmatrix}
H_{N-1} & H_{N-1} \\ 
 H_{N-1} & -H_{N-1}
\end{bmatrix}
\end{equation}

As in Haar, signal $x_{N}$ of length $N$ can be transformed as: $X_{N} = H_{N} \times x_{N}$. Similarly, 2-D Hadamard transform on a $N \times N$ signal, $y_{N}$ gives $Y_{N} = H_{N} \times y_{N} \times H_{N}^{T}$.

\section{\textbf{Preprocessing of data}}

\subsection{ (a) Ultrafast transient absorption spectroscopy}

\begin{figure}[h]
    \includegraphics[width=9cm]{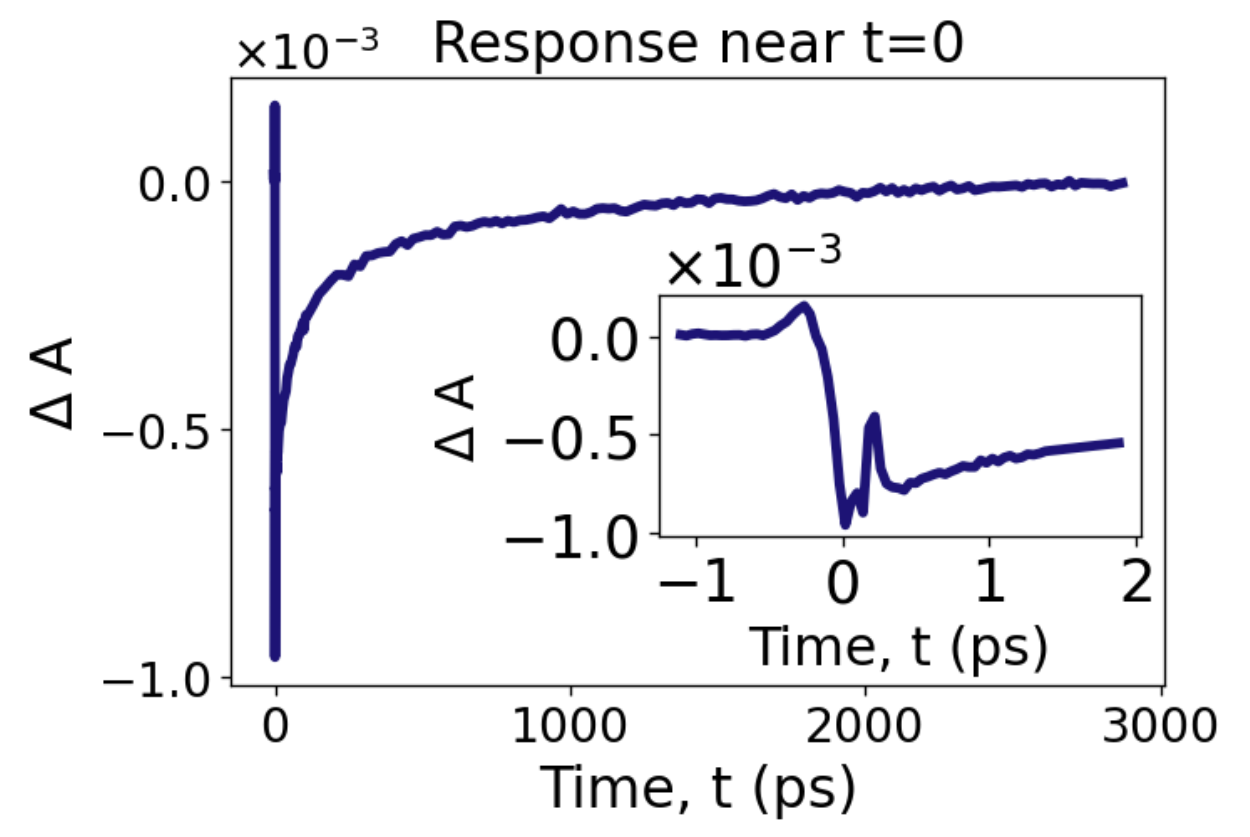}
    \caption{Transient absorption at 481 nm showing coherent, non-resonant response near $t=0$.}
    \label{fig:solvent_coherent}
\end{figure}

As we mentioned in the methods section that we excluded regions near $t=0$ for all analysis. Figure \ref{fig:solvent_coherent} shows transient absorption at $481$ nm. We see a response near $t=0$. Since, our specimen is a TiN nanoparticles in water, it is known that coherent, non-resonant response of the solvent can occur near $t=0$, hence we start our analysis from $t=0.78$ ps, avoiding this anomalous region.

Our initial data consists of 273 different wavelengths ranging from 421 nm to 743 nm, and each wavelength has $255$ absorption coefficients collected over time interval, $t=-5$ ps to $t = 2870$ ps. The 421 nm to 431 nm wavelength data are excluded. The data is very noisy in this wavelength range due to the very small amount of continuum photons generated by 800 nm light incident on the sapphire crystal and very low levels of light in the probe beam causes large digitization noise in the measured signal. After excluding this anomalous region, we interpolate the data such that each wavelength has 256 absorption coefficients.

\subsection{ (b) Ultrafast terahertz spectroscopy}
Our initial data set consists of 26 and 151  scan along $\tau$ and $t$ respectively. As before, we preprocess the data and interpolate it to make $32 \times 128$ for comparison among DCT, Haar and Hadamard. For the $\tau$ interval, we interpolate between the last two data points. For $t$ axis, we simply cutoff the data where the signal does not have any interesting feature. For the CS reconstruction, we randomly sample from $\tau$ and $t$. This corresponds to making a random coarse measurements in both $\tau$ and $t$ which significantly reduces the duration of experiment.

\section{\textbf{Example of sampling for ultrafast terahertz spectroscopy}}
 Here, we show an instance of what 15\% sampling looks like. We can see that more samples are concentrated near $\tau = 0$ because the time interval is very small compared to larger $\tau$ and lot more data is collected. 

\begin{figure} [h]
    \includegraphics[width=9cm]{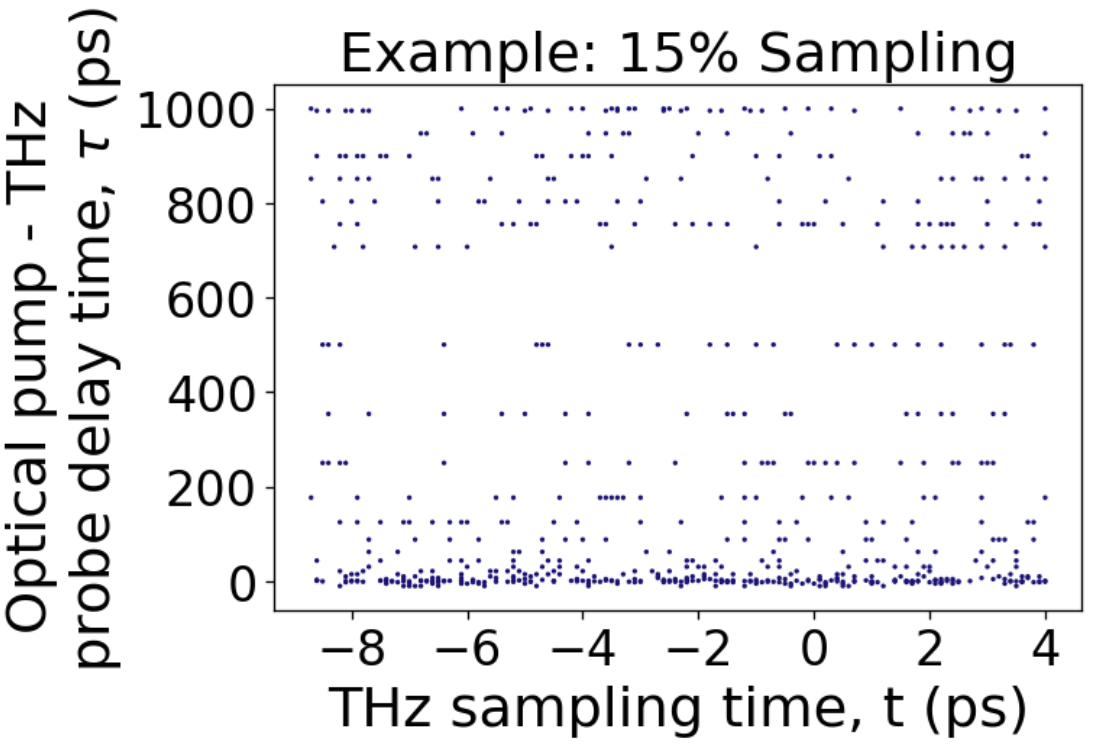}
    \caption{An instance of 15\% sampling. Most of the samples are concentrated near $\tau=0$ because the time interval is very short compared to larger $\tau$.}
    \label{fig:sampling}
\end{figure}

\section{\textbf{Cross validation method for choosing $\lambda$}}

\begin{figure}[ht]

    \includegraphics[width=18cm]{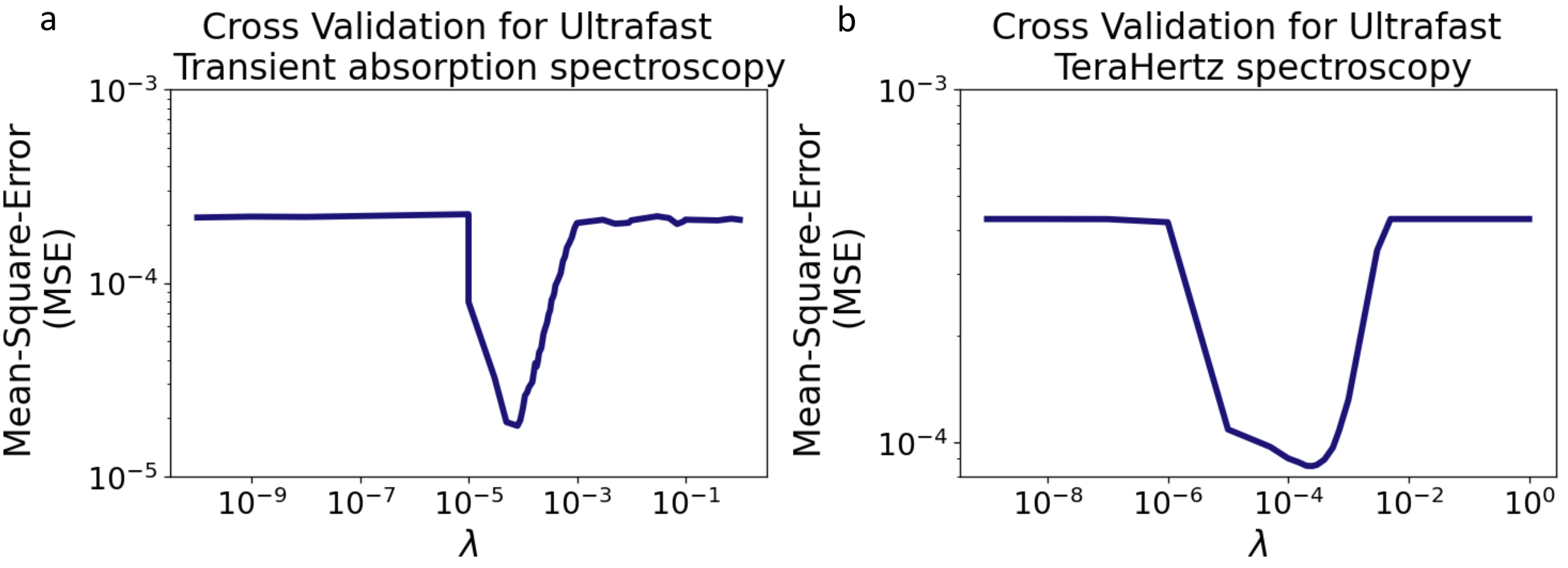}
    \caption{ (a) Cross-validation for choosing $\lambda$ for ultrafast transient absorption spectroscopy. The minimum MSE is given by $\lambda = 8 \times 10^{-5}$. (b) Cross-validation for choosing $\lambda$ for ultrafast terahertz  spectroscopy. The minimum MSE is given by $\lambda = 2 \times 10^{-4}$. }
    \label{fig:cross_validation}
    \vspace{-0.5cm}
\end{figure}

$\lambda$ is a free parameter which is data dependent. It is unknown beforehand what value of $\lambda$ to use for CS reconstruction. One approach to estimate $\lambda$ is by cross-validation which we use for both of our experimental data. For both experiments, we first randomly sample 20\% of the full data. We use 80\% of this sampled data for CS reconstruction and other 20\% for cross validation. The method work as follows: we sweep over a range of $\lambda$ values and reconstruct the signal. Then, for each value of $\lambda$, we calculate the mean-square-error (MSE) between reconstructed signal and the other 20\% data kept for cross validation. We then select the $\lambda$ value which gives the lowest MSE. For ultrafast transient absorption reconstruction, we chose the value of $\lambda$ that gives lowest MSE over all the wavelengths.  Figure \ref{fig:cross_validation}-a shows MSE as a function of $\lambda$ for ultrafast transient absorption. The value of $\lambda$ which gives minimum MSE is $8 \times 10^{-5}$, and we used this value for CS analysis of ultrafast transient absorption spectroscopy. Similarly, Figure \ref{fig:cross_validation}-b shows MSE as a function of $\lambda$ for ultrafast terahertz spectroscopy. The $\lambda$ value which minimizes the MSE is $2 \times 10^{-4}$ and we used this value for all CS analysis of ultrafast terahertz spectroscopy. 

\vspace{1cm}

\section{\textbf{Assessing quality of CS reconstruction}}

To access the quality of CS reconstruction, we looks at how at the absolute difference between the CS reconstruction and the full experiment. 

\subsection{(a) Ultrafast transient absorption spectroscopy}

\begin{figure} [h]
    \includegraphics[width=18.5cm]{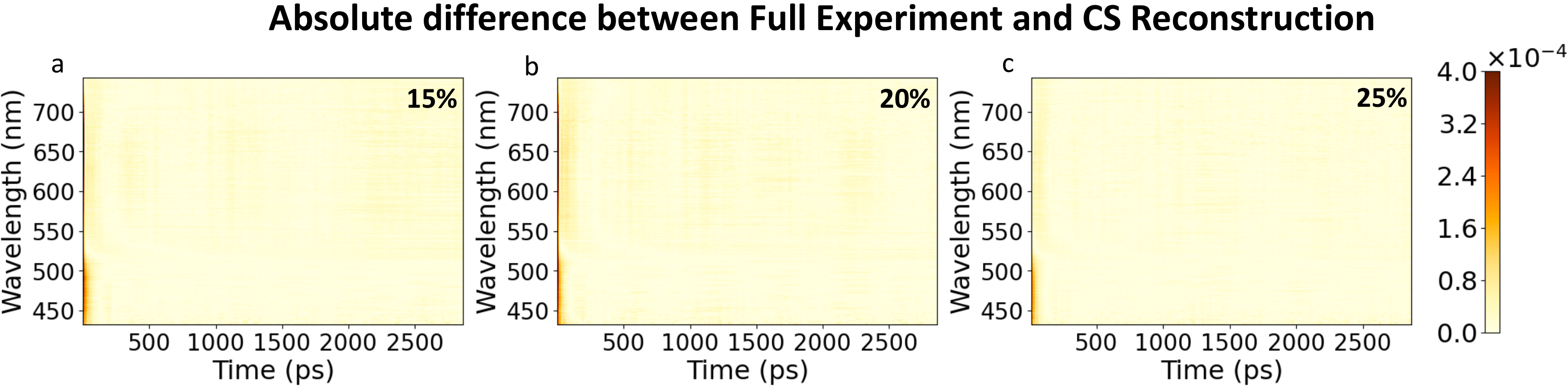}
    \caption{Figure shows the pixel-by-pixel magnitude difference between the full experiment and CS reconstructions at different sampling \%.}
    \label{fig:sampling}
    %\vspace{-.5cm}
\end{figure}

\vspace{1cm}
\subsection{(b) Ultrafast terahertz spectroscopy}

\begin{figure} [h]
    \includegraphics[width=18cm]{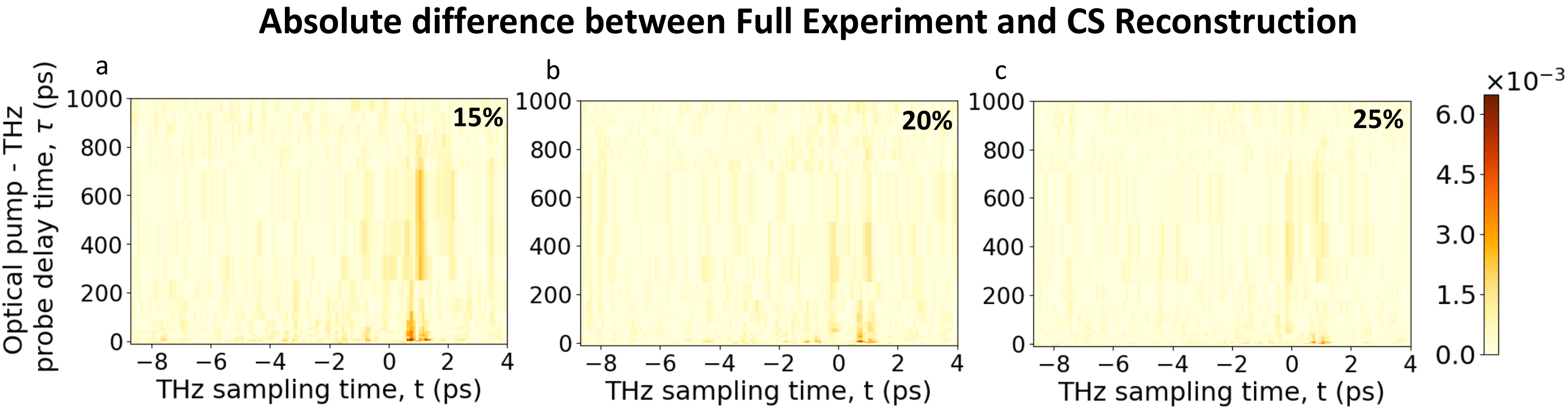}
    \caption{Figure shows the pixel-by-pixel magnitude difference between the full experiment and CS reconstructions at different sampling \%.}
    \label{fig:sampling}
\end{figure}

\section{\textbf{CS on another ultrafast terahertz spectroscopy experiment}}

\begin{figure} [h]
    \includegraphics[width=18cm]{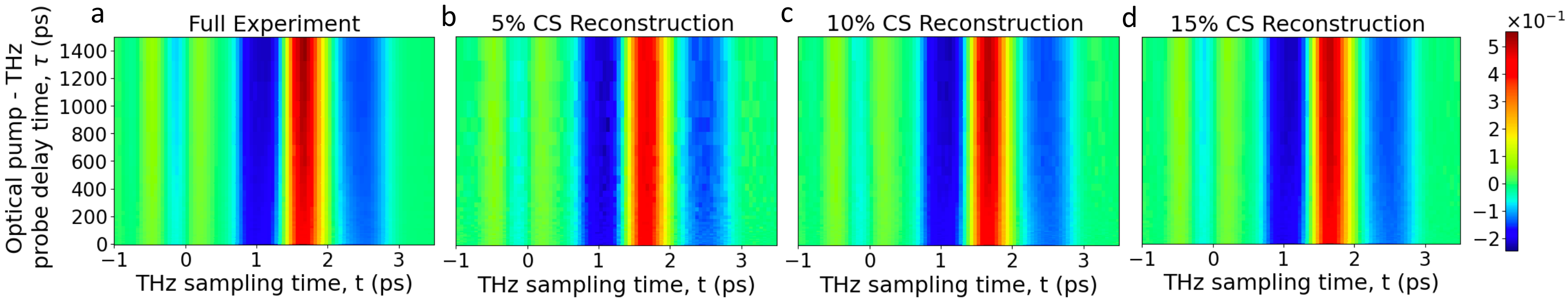}
    \caption{ (a) shows the full experiment. (b), (c) and (d) shows the CS reconstruction with 5\%, 10\% and 15\% samples, respectively}
    \label{fig:another_terahertz}
\end{figure}

We show another example of using CS in another ultrafast terahertz spectroscopy experiment. The specimen is a piece of silicon on an insulator (SOI) being optically pumped at about 650 nm near $\tau=0$ ps. We can see from Figure \ref{fig:another_terahertz}. that we can get fairly good estimate of the full experiment even with as low as 5\% sample.

\end{document}